\documentclass[aps,prl,twocolumn,
floatfix]{revtex4}
\usepackage{graphicx}
\usepackage{amssymb}
\usepackage{dcolumn}
\newcolumntype{.}{D{.}{.}{-1}}

\begin{document}

\title{Efficiency vs. multi-photon contribution test for quantum dots }

\author{Ana Predojevi\'{c}$^1$}
\author{Miroslav Je\v{z}ek$^2$}
\author{Tobias Huber$^1$}
\author{Harishankar Jayakumar$^1$}
\author{Thomas Kauten$^1$}
\author{Glenn S. Solomon$^3$}
\author{Radim Filip$^2$}
\author{Gregor Weihs$^1$}

\affiliation{$^1$Institute for Experimental Physics, University of Innsbruck, Technikerstrasse 25, 6020 Innsbruck, Austria}
\affiliation{$^2$Department of Optics, Palack\'y University, 17.~listopadu 12, 77146 Olomouc, Czech Republic}
\affiliation{$^3$Joint Quantum Institute, National Institute of Standards and Technology  and University of Maryland, Gaithersburg, MD 20849, USA}

\begin{abstract}
 The development of linear quantum computing within integrated circuits demands high quality semiconductor single photon sources. In particular, for a reliable single photon source it is not sufficient to have a low multi-photon component, but also to possess high efficiency. We investigate the photon statistics of the emission from a single quantum dot with a method that is able to sensitively detect the trade-off between the efficiency and the multi-photon contribution. Our measurements show, that the light emitted from the quantum dot when it is resonantly excited possess a very low multi-photon content. Additionally, we demonstrated, for the first time, the non-Gaussian nature of the quantum state emitted from a single quantum dot.
\end{abstract}


\maketitle

The ideal single photon state is quantum mechanically represented by the Fock state $|1\rangle$, a quantum counterpart of the classical particle. The particle nature of a single photon is traditionally verified by observing an anticorrelation effect on a beam splitter \cite{ Grangier86}. This measurement of the intensity autocorrelation is commonly accepted as the way to test a light source for non-classicality \cite{Scheel}. Nonetheless, such a measurement is an intensity-normalized measurement and therefore completely insensitive to the vacuum contribution, $|0\rangle$. In practice, however, all single-photon sources emit a stochastic mixture of the vacuum and the Fock state $|1\rangle$. For many applications, which do not depend on the efficiency of the single-photon source the intensity autocorrelation is a sufficient test. On the other hand, applications like linear optical quantum computing \cite {KLM} crucially depend on the overall source and detector efficiency \cite{Brien1,Varnava, Jennewein}.

Seen from this perspective, the characterization of a quantum state produced by realistic single photon sources would strongly benefit from a measurement that is sensitive to the vacuum contribution as well as the multiphoton component. By measuring the density matrix one obtains the full information about the quantum state of a real single photon source; regrettably, such a full state tomography is usually quite challenging. An alternative approach is to measure a quantum phase-space description of the state, its so-called Wigner function \cite{Leonhardt}. It is well known that the single photon state $|1\rangle$ has a negative Wigner function \cite{Glauber, Bachor}. Furthermore, there is a whole class of states that possess negative Wigner functions (so called non-Gaussian states of light) as shown by Hudson's theorem \cite{Hudson}. The Wigner function is usually measured by homodyne tomography. This is a very tedious procedure and in some cases even impossible.  However, when it is possible, it can detect the negativity of the states Wigner function, as was demonstrated for heralded single photon sources \cite{Lvovsky}. The practical obstacle for such a measurement is that the negativity depends crucially on the overall source and detector efficiency $\eta$. Namely, if $\eta<0.5$ the measured Wigner function is positive, even though the photon source may give a perfectly antibunched intensity autocorrelation measurement. Therefore, the Wigner function has never been reconstructed for solid-state single photon sources, mainly because the collection and detection efficiencies are very unfavourable.


Recently, some of us proposed a novel non-Gaussianity criterion (NG criterion) designed to characterize single photon sources \cite{Filip11}. The NG criterion is based on the measurement of photon statistics but, in contrast to the intensity autocorrelation measurement, is sensitive to the overall source (collection and detection) efficiency. Light characterization by measurement and reconstruction of the Wigner function shares some similarities with the NG criterion. In particular: it is sensitive to the losses and is able to distinguish Gaussian from non-Gaussian states \cite{Kimble, Schnabel, Akira}.  Additionally, this criterion \cite{Filip11} is still applicable in the case of low emission and detection efficiency in contrast to the direct measurement of the negativity of the Wigner function. In other words, this criterion enables an efficiency-sensitive evaluation of real single photon sources without the necessity for complete quantum state tomography.


In this letter, we present measurements performed on the light emitted by a single quantum dot and by a parametric down-conversion heralded single photon source. With these measurements we performed an advanced study of the statistics of the emission from a single quantum dot. In particular, we characterized the efficiency and multiphoton contribution and then tested the obtained results using the NG criterion \cite{Filip11}. Further, we compared the results obtained under different types of quantum dot excitation as detailed below. For reference, we also measured this on a parametric down-conversion-based heralded single photon source.



\begin{figure}[!b]
\centerline{\includegraphics[width=0.95\columnwidth]{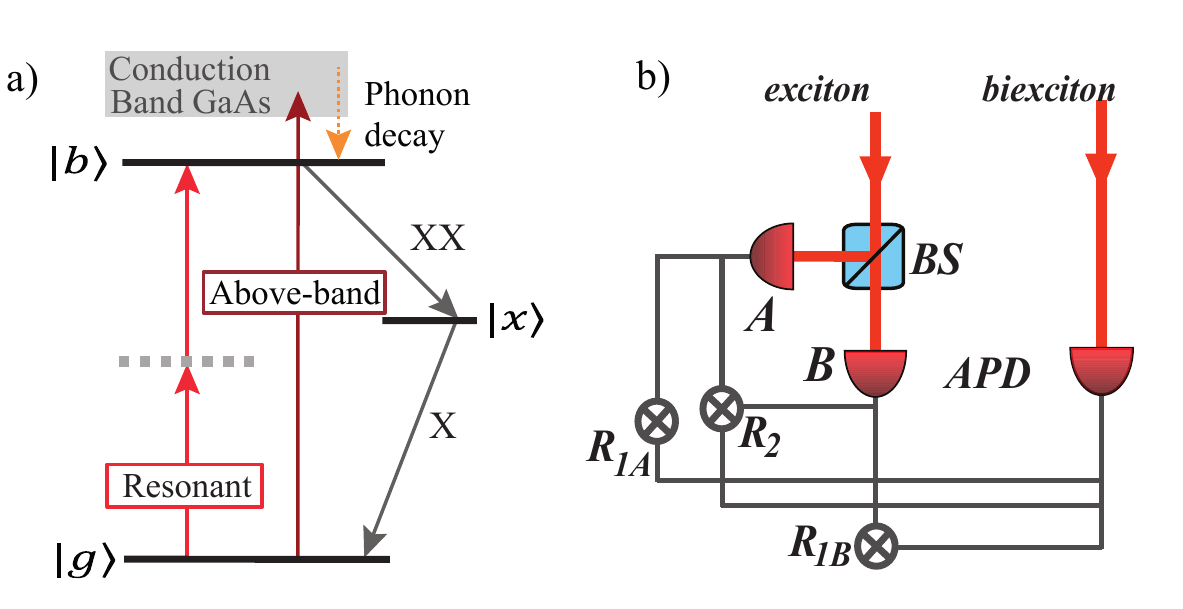}}
\caption{Excitation level scheme and detection scheme. a) Resonant excitation coherently drives
the two-photon transition between the ground $|g\rangle$ and the biexciton
$|b\rangle$ state via a virtual level shown as a dashed gray line. The system decays in a cascade via the exciton $|x\rangle$ state. Of the two possible decay paths we use only the vertical polarization.
Above-band excitation excites the carriers in the surrounding material. b) After spectrally resolving the emission on a diffraction grating (not shown in the figure) the spectral lines
of interest (exciton and biexciton) were separated and coupled into optical fibres. A fibre beamsplitter divided the exciton light onto two detectors for state verification. The biexciton detections were used as trigger events.}
\label{fig:schematic_qd2}
\end{figure}

To drive our quantum dot system we used two types of excitation: the first one continuous and above-band and the second one pulsed and resonant. In above-band excitation the energy of the excitation laser is much larger than the quantum dot emission energy. It produces carriers in the surrounding material that can be randomly trapped in the quantum dot potential. Secondly, we performed resonant excitation \cite{Jayakumar} by two-photon excitation of the biexciton state. Here, we exploited the biexciton binding energy \cite{Flissikowski,Jayakumar} in order to use an excitation laser not resonant with any single photon transition. By applying a laser pulse of a specific pulse area we drove the quantum dot system from the ground state to the biexciton state with very high probability (60$\%$) and in a coherent and controlled manner \cite{Jayakumar}.
The scheme of the energy levels of the quantum dot is shown in Fig.~\ref{fig:schematic_qd2}a and the detection scheme is shown in Fig.~\ref{fig:schematic_qd2}b.

The sample contained low density self-assembled InAs quantum dots embedded in a planar micro-cavity. It was placed in a continuous-flow cryostat and held at a temperature of 5~K. The excitation light was derived from a tunable Ti:Sapphire laser that could be operated in picosecond-pulsed (84~MHz repetition rate) or continuous-wave mode.

The light was directed onto a dot from the side, where we used the lateral wave-guiding mode of the micro-cavity \cite{Jayakumar}. The emission was then collected from the top. The exciton and biexciton photons were separated by a grating and sent into optical fibres. For the above-band measurements we used a single mode fibre for the biexciton spectral line and, in order to increase the detection efficiency, a multi-mode fibre and a beamsplitter for the exciton line. Under resonant excitation it was not possible to use multi-mode fibres due to the spectral proximity of the scattered laser so we used a single mode fibre for the biexciton line and a single mode fibre and a beamsplitter for the exciton line. Two single photon detectors detected the exciton photons and one detected biexciton photons (see Fig.~\ref{fig:schematic_qd2}b). The photon detection events were recorded by a multichannel event timer.

The NG criterion \cite{Filip11} uses two key parameters, the success rate (efficiency) $p_1$ and the error rate (multiphoton contribution) $p_{2+}$ of the single photon source, which directly apply to the use of single photon sources for linear optical quantum computing \cite{Jennewein}.

To analyse the photon statistics of the exciton emission triggered on the detection of a biexciton photon, we started by deriving the number of single counts, two-fold, and three-fold coincidences from the measured data, which includes the arrival times of all the recorded photon detection events. Here, single counts are detections of a biexciton photon (trigger event) without an exciton photon. A two-fold coincidence is the detection of a biexciton photon followed by a detection of an exciton photon in one of the arms of the beamsplitter. A three-fold coincidence stands for a detection event on all three detectors  (two exciton photons). The coincidence window was varied as detailed below.


We used the number of single counts, two-fold, and three-fold coincidences to estimate the contribution of vacuum $p_0$, single photon $p_1$, and multiphoton terms $p_{2+}$, to the emitted exciton signal, \cite{Filip11b}.
In particular, it is shown in  \cite{Filip11b} that the probability of no event, $p_0$, can be expressed as

\begin{equation}
p_0=1-\frac{R_{1A}+R_{1B}+R_2}{R_0}.
\label{p0est}
\end{equation}

Here $R_0$ is the total number of counts ("singles") in the biexciton signal. $R_{1A}$, $R_{1B}$ are total numbers of the two-fold coincidences between the biexciton signal and either exciton signals, respectively.  $R_2$  is the total number of the three-fold coincidences.
The lower bound estimator of the single photon contribution is given by

\begin{equation}
p_{1,est}=\frac{R_{1A}+R_{1B}}{R_{0}}- \frac{T^2+(1-T)^2}{2T(1-T)}\frac{R_{2}}{R_0},
\label{p1est}
\end{equation}

where $T$ is the splitting ratio of the beamsplitter used in the measurements. The contribution  $p_{2+}$ of the multi-photon terms can be estimated as
\begin{equation}
p_{2+}=1-p_0-p_{1}.
\label{p2plus}
\end{equation}

We determine the photon statistics ($p_0$, $p_1$, $p_{2+}$) using this method because we are interested in calculating $p_1$ as the lower bound estimator, $p_{1,est}$. The lower bound estimation takes into account the splitting ratio of the beamsplitter. In this way we avoid that the number of coincidences is artificially modified by an unbalanced beamsplitter.

\begin{figure}[!b]
\centerline{\includegraphics[width=0.9\columnwidth]{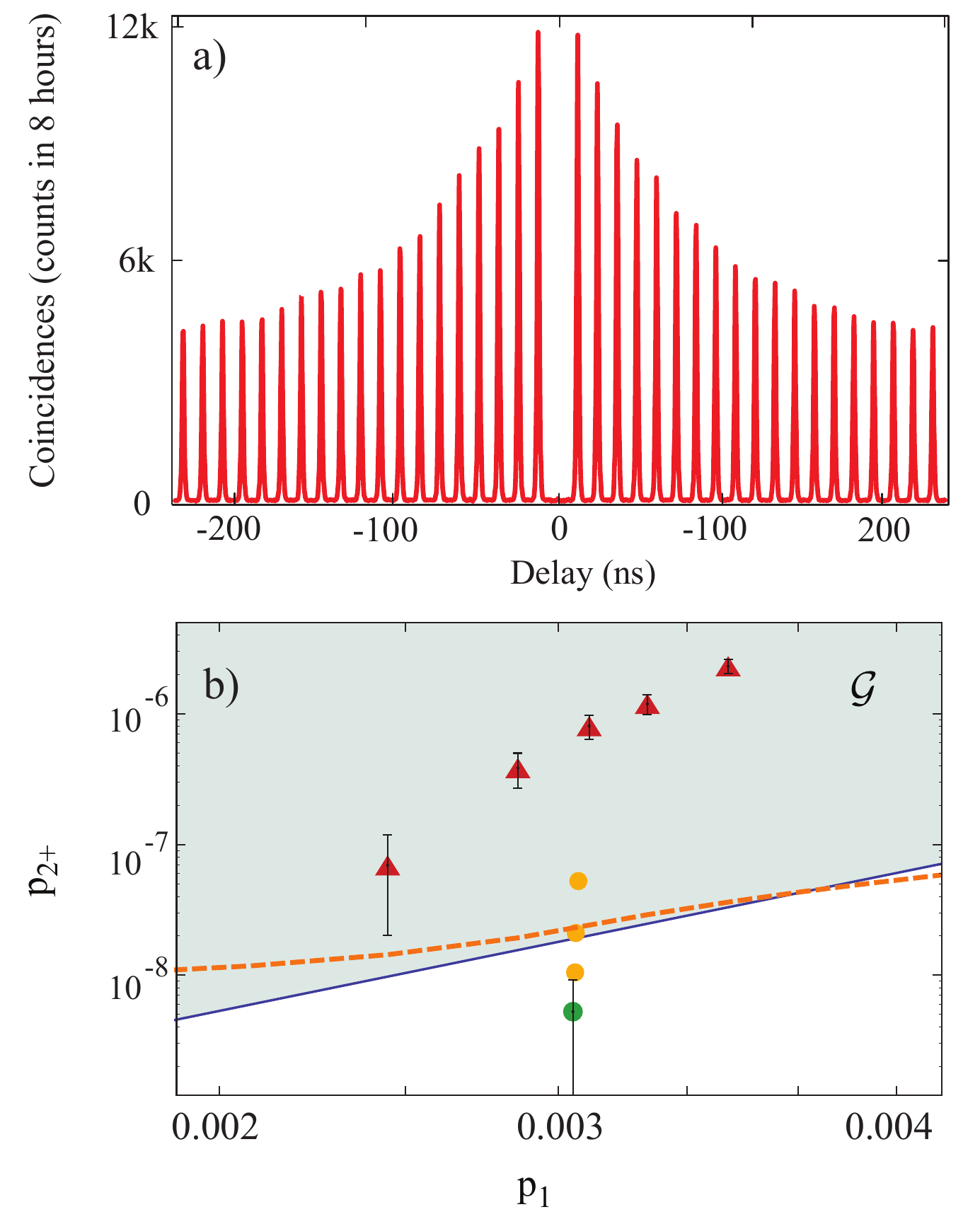}}
\caption{The intensity autocorrelation measurement and the multiphoton contribution, $p_2+$, plotted as a function
of the single photon contribution, $p_1$. a) The exciton signal shows excellent suppression of multi-photon events which can be
quantitatively expressed by intensity autocorrelation parameter of
0.0315(2). The plotted data was
acquired without the triggering on biexciton photon and is presented without
background subtraction. The decaying peak height observable on both sides of the graph results from the blinking of the  quantum dot \cite{Santori}. b) Here, $\mathcal{G}$
is the set of all mixtures of Gaussian states, and the lower, white region indicates non-Gaussian states. The circles stand for results obtained in resonant and pulsed excitation while
triangles for above-band and continuous wave excitation. In particular, the green circle stands for the result presented in the first row of the Table ~\ref{table2}, and the yellow circles for the results presented in the remaining rows. The error bars represent standard
deviations, the horizontal error bars of $p_1$ are smaller than the size of the symbols.
The solid blue curve represents the boundary presented in \cite{Filip11b} and given by Eq.~4. The orange dashed line marks the limit of the detection system in continuous excitation.}
\label{fig:h_log}
\end{figure}

The NG criterion defines a witness, $\Delta W$, that the given state is not a mixture of Gaussian states $\rho \notin \mathcal{G}$, where $\mathcal{G}$ is the set of all mixtures of Gaussian states \cite{Filip11b}.
It also derives a boundary (NG boundary) between the states that can be described as a mixture of Gaussian states and those that cannot. This boundary is given as \cite{Filip11b}
\begin{eqnarray}
p_0=\frac{e^{-d^2[1-\tanh(r)]}}{\cosh(r)}, \qquad
p_1=\frac{d^2 \, e^{-d^2[1-\tanh(r)]}}{\cosh^3(r)},
\label{p01bound}
\end{eqnarray}

where the squeezing constant $r$ \cite{Walls} is used to parametrize the curve with the displacement $d$ given by $d^2=(\mathrm e^{4r}-1)/4$ \cite{Filip11b}. The witness, $\Delta W$ is the directed minimum distance between the measured point ($p_0$, $p_{1,est}$) and the NG boundary. $\Delta W>0$ indicates that the measured state is non-Gaussian. The results are shown in Fig.~\ref{fig:h_log}b and Tables~\ref{table1} and~\ref{table2}.

Due to the very low collection and detection efficiency ( $\approx$ 0.3$\%$) the vacuum term $p_0$ prevails. The statistical uncertainties for $p_0$, $p_1$, and $p_2+$ were determined from the Poissonian statistics of the recorded events. In Tables~\ref{table1} and~\ref{table2} the last column gives the value of the witness, $\Delta W$, in units of the standard deviation.
For example, the result given in the first row of Table~\ref{table2} indicates a non-Gaussian state, (+), which is situated 2.63 standard deviations away from the boundary. In Fig.~\ref{fig:h_log}b, the states of light in the white area are non-Gaussian and produced by a source incompatible with only quadratic non-linear processes.

For the resonantly excited quantum dot our results unambiguously prove that the state cannot be expressed as a mixture of Gaussian states, because the witness, $\Delta W$ is positive for any coincidence window that is smaller than the repetition period of the laser pulses. For example, for a coincidence window of 10~ns the measured state exceeds the Gaussian boundary by 2.63 standard deviations (green circle in Fig.~\ref{fig:h_log}b). Extending the coincidence window to 10.24~ns, which includes the beginning of the consecutive pulse, we find the measured state to move towards the boundary of the Gaussian states. Further extension of the coincidence window places the state in the region where we cannot distinguish it from a coherent mixture of Gaussian states (shown in Fig.~\ref{fig:h_log}b in yellow circles).

\begin{table}[t]
\caption{Above-band excitation estimated probabilities $p_0$, $p_1$, $p_{2+}$
and the corresponding sign of the witness $\Delta W$  (given in the parenthesis in the last column) shown for several different
coincidence window widths $w$. The last column also indicates the distance of the measured point to the border separating the two classes of states. This distance is given in number of standard deviations,  $\sigma$  }
\label{table1}
\begin{ruledtabular}
\begin{tabular}{cccc@{}r@{\quad}}
\multicolumn{1}{c}{$w$\,[ns]}&
\multicolumn{1}{c}{$p_0$} &
\multicolumn{1}{c}{$p_1\,[\times10^{-3}]$} &
\multicolumn{1}{c}{$p_{2+}\,[\times10^{-8}]$} &
\multicolumn{1}{c}{$\Delta W$ [$\sigma$]}  \\ \hline
1.54  & 0.997553(6) & 2.446(6) &   $\hspace{1.4 mm}6.92 \pm 4.89$ &   \textrm{-} 1.21  \\
2.05  & 0.997140(7) & 2.859(7) &  $\hspace{1.4 mm}38.46 \pm 11.59 $ & \textrm{-} 3.18   \\
2.56 & 0.996885(7) & 3.114(7) &  $\hspace{1.4 mm}80.55 \pm 16.78 $ & \textrm{-} 4.67   \\
3.07 & 0.996660(7) & 3.339(7) & $119.20 \pm 20.40$ &  \textrm{-} 5.70   \\
3.84 & 0.996319(8) & 3.678(8) & $231.40 \pm 28.50$ &  \textrm{-} 8.00  \\
\end{tabular}
\end{ruledtabular}
\end{table}

\begin{table}[t]
\caption{Resonant pulsed excitation allows us to distinguish our state from
a mixture of Gaussian states, which is witnessed by $\Delta W>0$. As well as in the Table ~\ref{table1} The last column indicates the sign of the of the witness $\Delta W$, (+) indicating non-Gaussian and (-) Gaussian state. The distance is also given in number of standard deviations,  $\sigma$ }
\label{table2}
\begin{ruledtabular}
\begin{tabular}{cccr@{}r@{\quad}}
\multicolumn{1}{c}{$w$\,[ns]}&
\multicolumn{1}{c}{$p_0$} &
\multicolumn{1}{c}{$p_1\,[\times10^{-3}]$} &
\multicolumn{1}{c}{$p_{2+}\,[\times10^{-8}]$} &
\multicolumn{1}{c}{$\Delta W$ [$\sigma$}]  \\ \hline
10.00  & 0.996939(3)  & 3.061(3) & $0.52 \pm 0.52$ & + 2.63  \\
10.24  & 0.996938(3)  & 3.062(3) & $1.05 \pm 0.74$ & + 1.16  \\
10.75  & 0.996935(3)  & 3.064(3) & $2.10 \pm 1.05$ & \textendash \enspace   0.17  \\
11.24  & 0.996931(3)  & 3.067(3) & $2.62 \pm 1.17$ & \textendash  \enspace 0.60  \\
\end{tabular}
\end{ruledtabular}
\end{table}

The results obtained in above-band excitation (red triangle in Fig.~\ref{fig:h_log}) strongly depend on the chosen coincidence window. If the coincidence window is larger than the exciton lifetime, the dot may get re-excited and thus emit multiple photons. If on the other hand, the coincidence window is smaller than the lifetime, the efficiency $p_1$ will be reduced and eventually detector dark counts will dominate and form the noise floor. As a result for a decreasing coincidence window our data show a tendency of approaching the NG boundary, but cannot cross it. The overall measurement time in this case was close to 8 hours and the average single count rates in the heralding (biexciton) channel $S_0$ and the signal (exciton) channels $S_{1A}$ and $S_{1B}$ were \{5.4, 358, and 196\} kcounts per second, respectively. The beamsplitter ratio for these measurements was $T_{mm}$=0.64(1) and the detector dark count rate was 500~counts per second. Given these parameters we estimate a noise floor of 0.41 three-fold coincidences in eight hours for the smallest chosen coincidence window of 1.54~ns. This noise floor, which is depicted for the various coincidence windows as an orange dashed line in Fig.~\ref{fig:h_log}b), ultimately limits the sensitivity of the measurement system.

All the above-band excitation results shown in Fig.~\ref{fig:h_log} were calculated for coincidence windows longer (see Table~\ref{table1}) than the lifetime of the exciton state (0.71~ns). While we tried analyses using shorter coincidence windows we did not observe any three-fold coincidences then during the entire measurement run. This puts our estimation of $p_{2+}$ to zero, but will yield a statistical error that is much larger than the NG boundary itself providing only an inconclusive result. This exemplifies that our measurements and analysis do take into account the entire system of source, collection and detection and when we observe a non-Gaussian state it is a direct observation, not an extrapolation.

In comparison under resonant excitation the measured single rates were \{ $S_0$, $S_{1A}$ and $S_{1B}$\} = \{37, 20, and 17\} kcounts per second with the respective beamsplitter ratio of $T_{sm}$=0.54(1). The measurement time was 3 hours. The use of resonant excitation gives an excellent suppression of multiphoton events, as evident in Fig. \ref{fig:h_log}a. For resonant excitation we estimate the noise floor to be 0.04 three-fold coincidences per 8 hour measurement. This threshold is lower than the one we obtain in above-band excitation due to the different nature of the excitation. Namely, resonant excitation produces predominantly cascaded emissions therefore the heralding efficiency and thus $p_1$ is a bit higher than with above-band excitation despite the lower overall exciton detection rates.

It is also interesting to compare the results presented in Table~\ref{table2} with the traditional intensity autocorrelation measurement. We modified the post-processing of the measurements obtained in resonant excitation, and used the laser pulse arrival as the trigger event instead of the detection of the biexciton. Here, we obtain $p_{1}^{ac}$=0.43444(4)$\times10^{-3}$ and $p_{2+}^{ac}$=0.41(2)$\times10^{-8}$. The results show a multiphoton contribution that is comparable to the data presented in the first row of Table~\ref{table2}, but the efficiency of the source is significantly reduced. In particular, $p_{1}^{ac}$/$p_1$=0.14. This result is to be expected due to the blinking of the quantum dot and the imperfect emission probability achieved in resonant excitation. In \cite{Jayakumar} we showed, in measurements performed on the same single quantum dot, that about 2/3 of the time the blinking stops the emission from the device. Additionally, we demonstrated \cite{Jayakumar} that due to the level dephasing in the quantum dot the the maximal emission probability achievable is 70$\%$. The measurements presented here were obtained with a laser pulse energy capable of bringing 60$\%$ of the population from the ground to the biexciton state. In other words, we expected $p_{ac1}$/$p_1$=0.3$\times$0.6=0.18.
We can then can estimate $g_2(0)$ as 2$\times$[1-$p_0$-$p_1$]/[2$\times$(1-$p_0$)-$p_1$]$^{2}$ and the Grangier's \cite{ Grangier86} anticorrelation parameter $\alpha$=$R_0$$R_2$/$R_{1A}$$R_{1B}$, which is equal to $g_2(0)$ for a symmetric beamsplitter as in our case. For the biexciton triggered measurement in resonant excitation we obtain $g_2(0)$=0.0030(1). The same measurement triggered on the laser pulse arrival gives $g_2(0)$=0.041(2). The later result is comparable with the autocorrelation parameter extracted from the traditional intensity autocorrelation measurement of 0.031(2).

\begin{figure}[b]
\includegraphics[width=0.95\linewidth]{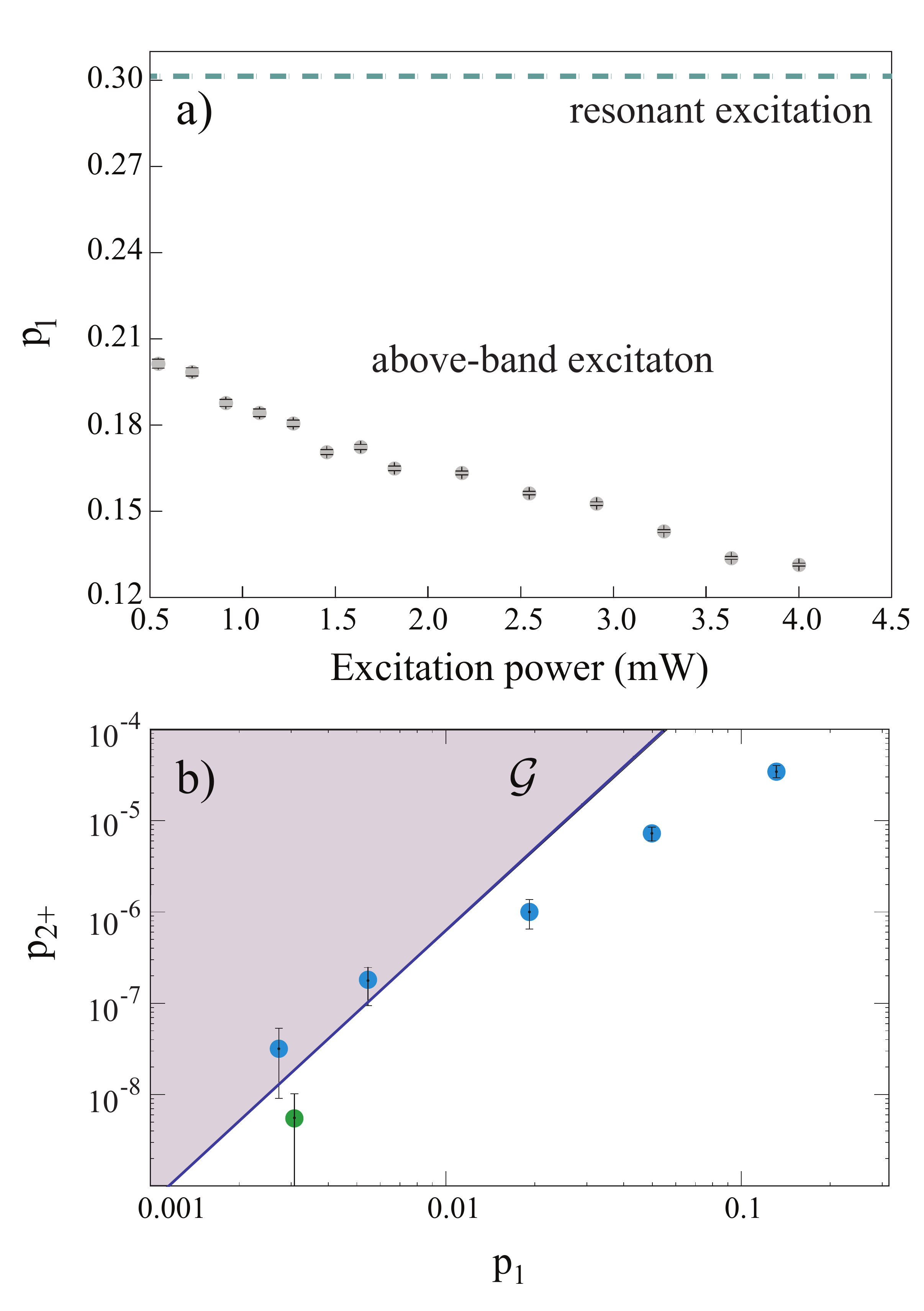}
\caption{Overall efficiency of the quantum dot photon source as a function of the excitation power and comparison between a single quantum dot and a down-conversion source. a) Here, the blue dashed line marks the probability of the detection of the single photon from a quantum dot under resonant excitation, $p_1$. The gray circles show the same probability under above-band excitation. For the latter we varied  the excitation power up to the saturation of the biexciton (4~mW) and observe a decrease of $p_1$. All measurements presented in this figure were obtained using single mode fibres to collect the quantum dot emission. The coincidence window for these data was 7~ns. b) The blue dots are results of measurements performed on the emission for a down-conversion source. Here, $p_1$ is gradually reduced through attenuation. The green dot shows the result for the quantum dot. The same point is plotted, also in green, in Fig~\ref{fig:h_log}.}
\label{fig:sagnac}
\end{figure}

As stated above, due to the different nature of the excitation, resonant and above-band, we obtain different noise thresholds. In addition, it is expected that the efficiency of these two processes is different. Only a coherent process such as resonant excitation can bring 100$\%$ of the population from the ground to the excited state. To compare the respective efficiencies we performed biexciton-triggered measurements of $p_1$ in above-band excitation for various excitation powers. The results are shown as gray dots in Fig.~\ref{fig:sagnac}a. For comparison the dashed line shows the $p_1$ value we obtained in resonant excitation. An interesting feature is that $p_1$ decreases with excitation power, reaching its minimum at the biexciton line saturates. We believe that this happens because the biexciton is re-excited directly from the exciton state before the system has reached the ground state and thus couldn't emit an exciton photon. Therefore the ratio of emitted exciton to biexciton photons is reduced. This is ultimately limits how strongly quantum dot-based pair sources can be driven incoherently.

To complete the study we also measured the photon statistics from a parametric down-conversion heralded single photon source. For this purpose we used the Sagnac-interferometer-based down-conversion source of entangled photon pairs described in \cite{sagnac}. The pair production rate was kept at the low value of 0.003 pairs per pulse in order to maintain a high quality of the entanglement \cite{comment1}. It has been shown that entanglement decreases with increased pump power and photon pair creation probability \cite{kuzucu}. The results are given in Table ~\ref{table3} and Fig.~\ref{fig:sagnac}b. Here, we used the signal photons to trigger the measurement (corresponding to the biexciton photons) and the idler photons were sent onto a fibre beamsplitter (corresponding to the exciton photons).

A quantum dot is a point like source in a large refractive index medium that emits light in all directions. This limits the collection efficiency into a single mode fibre to 1.5$\%$ in our case.  On the other hand, the photons produced in down-conversion are well directed in space and we can collect them into a single mode fibre with 74$\%$ efficiency. Furthermore, the quantum dot emission has a wavelength of around 920~nm while our down-conversion source produces photons at 808~nm. This yields detector quantum efficiencies of 0.2 and 0.5, respectively. The initial $p_1$ of the down-conversion source was 14$\%$ and in the experiment we gradually attenuated the idler beam to simulate losses comparable to the quantum dot case. The results show that for comparable overall efficiency and for the range of operating parameters available, $p_1$ the down-conversion source emission contains a higher proportion of multiphoton events than the emission from a single quantum dot.

\begin{table}[t]
\caption{Photon statistics measurement was performed on a down-conversion source. The coincidence window was here equal for all the measurements $w$=1.2~ns }
\label{table3}
\begin{ruledtabular}
\begin{tabular}{rrcl@{}r@{\quad}}
\multicolumn{1}{c}{$p_0$} &
\multicolumn{1}{c}{$p_1\,[\times10^{-3}]$} &
\multicolumn{1}{c}{$p_{2+}\,[\times10^{-8}]$} &
\multicolumn{1}{c}{$\Delta W$ [$\sigma$]}  \\ \hline
0.8685(2)  & 131.4(3) & $3477 \pm 941$ &  + 146  \\
0.95018(7)  & 49.81(7) & $725.4 \pm 123$ & + 56.7  \\
0.98081(3)  & 19.18(3) & $100.7 \pm 35.6$ & + 10.1  \\
0.99455(1)  & 5.45(1) & $19.20 \pm 8.59$ & \textendash \enspace 0.98  \\
0.997277(5)  & 2.723(5) & $3.11 \pm 2.20$ & \textendash \enspace 0.80  \\
\end{tabular}
\end{ruledtabular}
\end{table}

In conclusion, we demonstrated the non-Gaussian nature of the emission of a single quantum dot under resonant excitation.  With this we detected a higher order non-classicality than usually detected by autocorrelation measurements. Therefore, we gained an intrinsically  higher sensitivity to possible contributions from other emitters \cite{Filip11}. In particular, we used a pulsed laser to resonantly bring the quantum dot system from the ground to the biexciton state which showed exceptionally pure quantum states of light.

Our measurement is the first demonstration of the non-Gaussian nature of photons produced by a semiconductor device. For completeness, we contrasted our results with the traditional auto-correlation measurement and with a parametric down-conversion heralded single photon source. We concluded that for a comparable overall efficiency quantum dot single photon source shows a smaller multiphoton contribution.

The non-Gaussian nature of a quantum state is a very important resource for quantum communication \cite{Eisert} and quantum computing \cite{Niset, Veitch, Mari}. Furthermore, it is a fundamental property of the single photons state \cite{Glauber}. With the increased detection efficiencies available in other device geometries \cite{Pascale} the criterion we used becomes an important measure of the quality of the produced light. The combination of resonant excitation, deterministic excitation of the quantum dot, and very pure single photon states is essential for using semiconductor photon sources for integrated linear optical quantum computing \cite{KLM,Brien, Brien1,Kok}.

This work was funded by the European Research Council (project EnSeNa) and
the Canadian Institute for Advanced Research through its Quantum Information
Processing program. G.S.S. acknowledges partial support through the Physics Frontier Center at the Joint Quantum Institute (PFC@JQI). R.F. acknowledges project P205/12/0577 of GA\v{C}R.


\begin{thebibliography}{99}
\bibitem{comment1}
At production rate of 0.003 pairs per pulse we measured H/V, and A/D visibilities to be 99.86(3)$\%$ and 98.65(3)$\%$, respectively. The corresponding value of the tangle was T=0.961(7).
\end{thebibliography}
\end{document}